[1994/12/01]
\documentclass{zamm}
\usepackage{amsmath,amssymb,amscd}
\usepackage{epsfig}
\makeatletter
\DeclareRobustCommand{\BibTeX}{B\kern-.05em%
     \hbox{$\m@th$
           \csname S@\f@size\endcsname
           \fontsize\sf@size\z@
           \math@fontsfalse\selectfont
           I\kern-.025emB}%
     \kern-.08em%
     \-\TeX}
\makeatother

\newcommand{\reff}[1]{{\rm (\ref{#1})}}
\newcommand{\OO}{{\cal O}}
\newcommand{\be}{\begin{equation}}
\newcommand{\BENN}{\begin{equation}\nonumber}
\newcommand{\ee}{\end{equation}}

\newcommand{\real}{{\mathbb R}}
\renewcommand{\natural}{{\mathbb N}}
\newcommand{\integer}{{\mathbb Z}}
\newcommand{\complex}{{\mathbb C}}

\newcommand{\cR}{{\cal R}}
\newcommand{\eps}{{\varepsilon}}
\renewcommand{\epsilon}{{\varepsilon}}

\newcommand{\rand}[1]{}
\newcommand{\R}{\real}

\newcommand{\citel}[1]{\rand{#1}\cite{#1}}

\newcommand{\ba}{\begin{array}}
\newcommand{\ea}{\end{array}}

\newcommand{\CC}{{\cal{C}}}
\let\phi=\varphi
\newcommand{\Hhh}{{\rm H}^2_2}
\renewcommand{\L}{{\rm L}}
\newcommand{\Umf}{U_{{\rm mf},[q,\epsilon,c]}}
\newcommand{\Umfnn}{U_{{\rm mf},[0,\epsilon,c ]}}
\newcommand{\Utv}{U_{{\rm TV},[q,\epsilon ]}}
\newcommand{\Utvnull}{U_{{\rm TV},[0,\epsilon ]}}
\newcommand{\Utvfirst}{U_{{\rm TV},[q,\epsilon,\vartheta ]}}
\newcommand{\uper}{u_{{\rm per},[q,\epsilon ]}}
\newcommand{\vlim}{v_*}
\newcommand{\Vlim}{V_*}
\newcommand{\phistar}{\phi_*}
\newcommand{\Aeq}{A_{{\rm eq},[q,\vartheta]}}
\newcommand{\Aeqnull}{A_{{\rm eq},[0,0]}}
\newcommand{\Bfront}{B_{{\rm f},[c]}}
\newcommand{\vf}{v_{{\rm f},[c]}}
\newcommand{\hhU}{U_{\kc}}
\newcommand{\kc}{k_{\rm c}}
\begin{document}

\title[Nonlinear stability of bifurcating front
solutions]
{Nonlinear Stability of Bifurcating Front
Solutions for the Taylor-Couette Problem}

\subjclass{35Q30, 76E99}
\keywords{modulated fronts, diffusive behavior}

%
%

\Received{}

\Volume{78}
\Number{0}
\Year{2000}
\beginpage{1}


\author{Jean-Pierre Eckmann}
\author{Guido Schneider}
\authoraddress{Prof.~{\sc Jean-Pierre Eckmann};
D\'epartment de Physique Th\'eorique,
Universit\'e de Gen\`eve,
CH-1211 Geneva 4,
Switzerland}
\authoraddress{Prof.~{\sc Guido Schneider}; Mathematisches Institut,
Universit\"at Bayreuth,
D-95440 Bayreuth,
Germany}

\begin{abstract}
We consider
the Taylor-Couette problem
in an infinitely extended cylindrical domain.
There exist modulated front solutions which describe the
spreading of
the stable Taylor vortices into the region of the
unstable Couette flow.
These transient solutions have the form of a
front-like envelope advancing in the laboratory frame and leaving
behind
the  stationary,
spatially periodic Taylor vortices.
We prove the nonlinear stability
of these solutions
with respect to
small spatially localized perturbations.
\end{abstract}

\begin{acknowledgement}
Guido Schneider would like to thank for
the kind hospitality at the Physics
Department  of the  University of Geneva. This work is partially
supported by the Fonds National Suisse.
The work of  Guido Schneider is partially supported  by the
Deutsche Forschungsgemeinschaft DFG
under the grant Mi459/2--3.
\end{acknowledgement}

\maketitle

\section{Introduction}
Ludwig Prandtl's boundary layer theory of 1904 was a breakthrough
in modern fluid mechanics. It allows to distinguish two domains for
small viscosity flows.
In a narrow neighborhood of the boundaries viscous effects
play a role while outside this boundary layer the flow can be considered
as a potential flow.
The Taylor-Couette problem we consider here is a
hydrodynamic problem where the onset of instability
is due to the motion of the boundaries.
While Prandtl's ideas are still
relevant, they do not apply as such
in our case because the infinite length of the boundaries
will
produce an infinite amount of energy.

The Taylor-Couette problem consists in describing the
flow of a viscous, incompressible fluid in the domain
between two (counter-) rotating cylinders. More precisely,
we are interested in the velocity field of
a viscous incompressible fluid filling  the domain
$ \Omega=\R \times \Sigma $ between two concentric rotating
infinite cylinders, where $ \Sigma \subset \R^2 $
denotes the bounded cross section.
The flow between the rotating cylinders
is described by the Navier-Stokes equations
in $ \Omega $ with no-slip boundary conditions.
For simplicity we assume throughout this paper
the outer cylinder to be at rest.

When the rotational 
velocity of the inner cylinder is small, a stationary flow appears,
called the Couette flow $ U_{\rm Cou} $
with associated pressure field $ p_{\rm Cou} $.
It is homogeneous along the cylinders and the streamlines
are given by concentric circles.
For small Reynolds number $ \cR $ (which is proportional to the rotational
velocity of the inner cylinder)
these solutions are  exponentially stable.

A perturbation $(U,p)$ of the Couette
flow satisfies the
Navier-Stokes equations
\begin{eqnarray}
\label{ns1}
\partial_t U & = & \Delta U
- \cR [(U_{\rm Cou} \cdot \nabla)U+  (U \cdot \nabla) U_{\rm Cou}
+(U \cdot \nabla)U]
 - \nabla p~, \\ \label{ns2}
\nabla \cdot U & = & 0~,
\end{eqnarray}
with Dirichlet boundary conditions
\begin{equation} \label{ns3}
  U|_{\R \times \partial \Sigma}  = 0~.
\end{equation}
This problem has a unique solution
$ U $ and $ \nabla p $
if we add the vanishing mean flux condition,
\begin{equation} \label{ns4}
 [U_{({\rm x})}]_{\Sigma}(x)=\frac{1}{|\Sigma|}
\int_{z \in \Sigma} U_{({\rm x})}(x,z) dz =0~,\quad \text{for all } x\in\real~,
\end{equation}
where $ U_{({\rm x})}  $ stands for the velocity component along the infinite
$ x $--direction. See \cite{CI94}.

The trivial branch of solutions of the system \reff{ns1}--\reff{ns4}
is $U(x,z)\equiv 0$, which corresponds to the Couette flow.
It becomes unstable when
the Reynolds number $ \cR $ goes beyond a certain threshold
$ \cR_c $.
We define  $ \epsilon>0$ by $\epsilon ^2 = \cR - \cR_c $ and assume
throughout $\epsilon \ll1$.

For $\epsilon > 0$ a family of small spatially periodic equilibria
(Taylor vortices) bifurcates from the trivial solution $U\equiv0$:
\begin{equation}
\label{star}
\Utvfirst(x,z)  = 
\epsilon \ A \sqrt{1- c_1 q^2}
\ 
\hhU(z) e^{i (\kc + \epsilon q) (x+\vartheta)}+ \mbox{c.c.}+
 \OO(\epsilon^{2}),\qquad
  q \in [0,1/\sqrt{c_1}), \ \vartheta \in \real~.
\end{equation}
In \reff{star},
$ c_1$ is a positive constant, $ \kc  > 0 $ is a critical wavenumber, and
$ \hhU(z) \in \complex^3 $.
We will henceforth assume $\vartheta=0$ and omit it from the parameters of
$U_{\rm TV}$.
Such solutions have been constructed in
\cite{Ki66,OI68,KS69}.
It turned out that the linearization around the
stationary solutions $ \Utv $ possesses
continuous spectrum at least up to the imaginary axis.
These solutions
are linearly stable for $ c_1 q^2 \leq
\frac{1}{3}$ and so-called sideband- or Eckhaus unstable for $ c_1 q^2 >
\frac{1}{3}$. See   \cite{Eck65} or \cite{CE90a}.

In actual experiments,
the transition from the Couette flow to the Taylor vortices usually
begins from perturbations at the ends of the (finite) cylinders. 
A first vortex forms near the end of the cylinder, and from this seed
new vortices form away from the ends, and an advancing front seems to
move through the cylinder, leaving stationary, spatially periodic
vortices behind. Finally the cylinder is filled with Taylor vortices.

These advancing fronts cannot just be described as $U(x,t) = U_{\rm f}(x-ct)$,
because of the pattern they leave in the laboratory frame.
Rather
 they are of
the form described in
\begin{theorem}
\cite{HS99}. For each $q$ with $c_1 q^2 < \frac{1}{3}$
the following holds.  For
sufficiently small $\epsilon >0 $  and for $c>0$
the system \reff{ns1}--\reff{ns4}
 possesses
modulated front
solutions
\BENN
U(x,z,t) = \Umf (x- ct,x,z)~,\end{equation}
with
\BENN \Umf(\xi,x,z)
=\Umf (\xi,x+ \frac{2\pi}{\kc +\epsilon q},z)~,
\end{equation}
and the boundary conditions
\begin{eqnarray*}
\lim_{\xi \to -\infty}
\Umf(\xi,x,z) & = 
\Utv(x,z)~,\quad & {\text{the Taylor vortices}}~,  \\
 \lim_{\xi \to \infty}\Umf (\xi,x,z) &\, \,\equiv \,\,0~,\quad & {\text {the Couette flow}}~.
\end{eqnarray*}
\end{theorem}
\begin{figure}[ht]
\input pra3lab.tex
\epsfig{file=maple106.eps, width=16cm,height=4cm, angle=0}
\vspace*{-4.15cm}

\epsfig{file=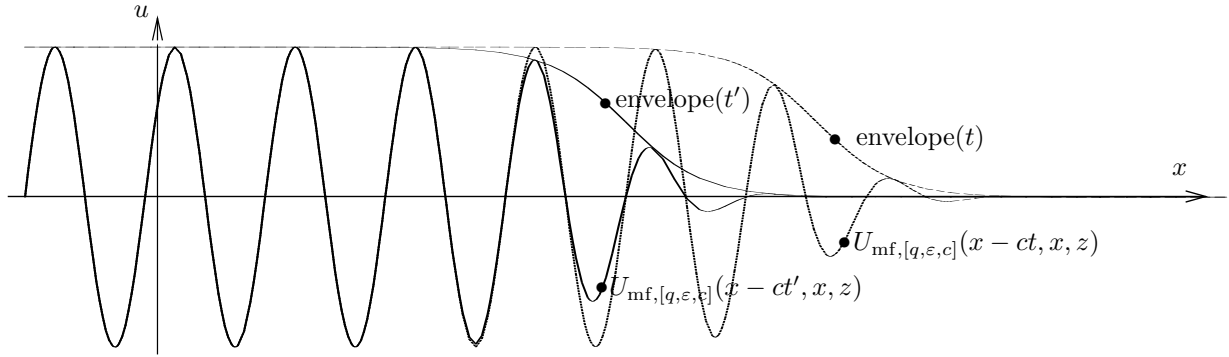, width=16cm,height=4cm, angle=0}
\caption{{Sketch of a modulated front, advancing from left to right at
times $t'<t$, and the advancing envelopes.}}
\end{figure}
In the present paper we prove the nonlinear stability---under
spatially localized 
perturbations---of these
bifurcated front solutions $\Umf$.
In particular, this means that a small perturbation cannot
destroy the growing regular pattern in the bulk, nor make it
move. This is not an obvious result since there could be a 
soft mode in the system, due to translation invariance. But, although
the Taylor vortices are only diffusively stable, their positions are pinned.
The precise statement of the result
can be found in Theorem \ref{thhaupt}.

Our proof is an adaptation of our previous paper
\cite{ES00}
where we dealt with the stability of front solutions
for the Swift-Hohenberg equation
\BENN
\partial_t u = - (1+ \partial_x^2)^2 u + \epsilon^2 u - u^3~,
\qquad u(x,t) \in \real, \ x \in \real, \ t \geq 0  
\end{equation}
connecting
the stable periodic equilibria
\BENN 
\uper = \eps \sqrt{\frac{1-4 q^2}{3}}
e^{i(1+ \eps q)x} + {\rm c.c.} + \OO(\epsilon^2)~,  \end{equation}
for $ 4 q^2 < 1/3 $
with the unstable trivial solution $ u = 0 $.

The stability proof is based on the following idea.
Consider a spatially localized perturbation
ahead of the front. In this unstable region any perturbation will
grow exponentially because 0 is an unstable solution.
But, after a {\it finite} time the front will have
reached the
perturbation, or equivalently, the perturbation has entered
the stable region  behind the front. This region is
{\it diffusively stable} and thus perturbations decay as
$ 1/\sqrt{t} $. Therefore, one expects the front to be stable under
sufficiently small spatially localized perturbations
which decay exponentially with the
distance ahead of  the front.
In particular, the front is stable for sufficiently small
perturbations with compact support.

In order to make these ideas clearer,
we describe first two easier model problems.
In Section \ref{sec2} we explain for  a very simplified model
how 
weighted norms  can be used to stabilize the
a priori unstable region ahead of the front.
In this very simplified model the equilibrium in the bulk will be
assumed
to be stable with some {\it exponential} rate.
In Section \ref{sec2gl}
we give up the exponential stability of the  equilibrium
and replace it by a
diffusively stable equilibrium, i.e., the continuous
spectrum of the linearization now reaches the imaginary axis
and is no longer separated from it.
As an example we consider the amplitude equation associated with
the Taylor-Couette problem, namely the Ginzburg-Landau equation.
For this equation it will be more or less obvious that
spatially localized perturbations of
the
stable equilibrium decay as a solution of a linear diffusion equation.
For the Taylor-Couette problem,
this is no longer obvious for the solutions we are interested in and
so
the main difficulty
for the fronts $ U_{\rm mf} $
is to recover the diffusive behavior behind
the front.
\begin{remark}
Since
the internal
wavelength of the spatially  periodic flow is in general much smaller
than the length of the cylinders and due to the fact that
the ends of the cylinders
in a laboratory experiment
do not influence the form of the pattern in the interior,
we consider the Taylor-Couette problem
with cylinders of infinite length.
As explained, up to a certain extent this idealization
is a good description of Nature.
\end{remark}
\begin{remark}
The Taylor-Couette problem is one of several examples
where our theory applies.
Other examples are B\'enard's problem and reaction-diffusion systems.
Loosely speaking our theory applies whenever a spatially homogeneous state
becomes unstable via a supercritical
stationary bifurcation at a non-zero Fourier wavenumber
in a translationally invariant system.
\end{remark}
\begin{remark}
The existence proof for modulated fronts
$ U(x,z,t)=\Umf (x- ct,x,z)$
connecting the stable Taylor vortices with the
Couette flow
is based on center manifold theory for elliptic problems
on unbounded cylindrical domains
\cite{Ki82}. It is complicated here by an infinite
number of eigenvalues on the imaginary axis for $ \cR = \cR_c $.
The related result for the Swift-Hohenberg equation can be found in
\cite{CE86,EW91}.
\end{remark}

\section{Basic ideas}
\label{sec2}

The ordinary differential equation
$
\partial_t u = u $ with $ u(t) \in \R $
is used to model the exponential growth of a certain quantity $ u $,
for instance
a chemical concentration  or a
population of bacteria.
Usually
nonlinear saturation  effects
give some bound on the possible  size of $ u $.
This can be taken into account by adding a nonlinear term, for example
$ -u^2 $,
to the model, leading to
\begin{equation} \label{a2}
\partial_t u = u - u^2~.
\end{equation}
This nonlinear model
possesses in addition to the  unstable equilibrium  $ u = 0 $ also
a stable equilibrium $ u=1 $, and so every solution with initial
condition in the interval $ (0,1) $ converges  towards
$ u = 1 $ for $ t \rightarrow \infty $.
This model is refined further by taking into account the spatial structure
of the solutions:
Chemical reactions or the growth of a population in general
do not happen in a spatially homogeneous manner and so the above model is
extended by adding a diffusion term $ \partial_x^2 u $ to
the system, i.e.,
\begin{equation}
\label{a3}
\partial_t u = \partial_x^2 u + u - u^2~,
\end{equation}
with $ u(x,t) \in \R $.
For solutions with small positive spatially localized initial conditions
the dynamics is very simple.
Solutions with such initial conditions develop in a universal way.
By the reaction term $ u -u^2 $ at fixed $ x \in \R $
 the solutions are  drawn to the stable equilibrium
$ u = 1 $. Then,
by the diffusion term the part of the solution which is close to
the stable phase $ u =1 $ starts to  spread into the region
of the unstable phase $ u = 0 $ forming two fronts
with velocities  $ c =\pm 2 $, one to
left and one to the right.

\begin{figure}[ht]\label{f2}
\input pra1lab.tex
\epsfig{file=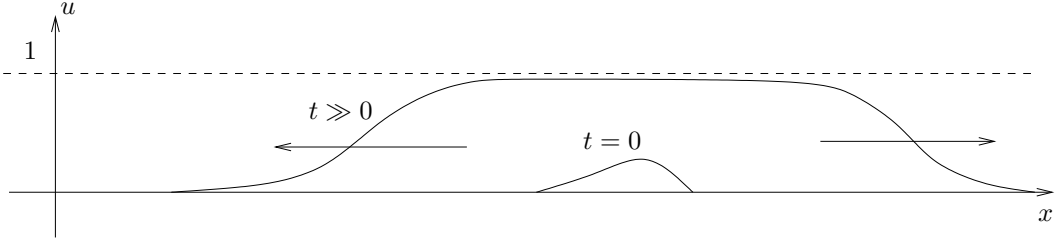, width=14cm,height=3cm, angle=0}
\caption{{ Two fronts for \reff{a3},
generated by localized initial data.}}
\end{figure}

Such front solutions $ u(x,t) = \vf(x-ct) = \vf(\xi) $
connecting the stable phase $ u = 1 $ for $ \xi \rightarrow - \infty $
with the unstable phase $ u = 0 $ for $ \xi \rightarrow  \infty $
can be found by
analyzing the second order
ordinary differential equation satisfied by $ v=\vf $
in the $ (v,v') $--phase plane. 
The partial differential
equation \reff{a3} can be considered in the  moving frame $ \xi = x- ct $,
where it takes the form
\begin{equation}\label{moving}
\partial_t v = \partial_{\xi}^2 v + c \partial_{\xi} v + v - v^2~.
\end{equation}
A front $\vf$ then satisfies $\vf''+c\vf'+\vf-\vf^2=0$.
It can be found by showing that there exists a heteroclinic connection
between $ (v,v') = (1,0) $ and $ (v,v') = (0,0) $, where the latter
fixed point has real eigenvalues when $ |c| \geq 2 $.
As a consequence, when $ |c | \geq 2 $ the fronts satisfy
$  \vf(\xi) \in (0,1) $.

Thinking of a chemical reaction front it is obvious that the stability
of these fronts is a relevant question.
We recall that an equilibrium $ y =0 $ of an
ordinary differential equation $ \dot{y} = A y + \OO(y^2) $
is called (Lyapunov) stable in $ \R^d $ if for all $ \epsilon> 0 $ there exists a
$ \delta > 0 $ such that $ \| y(0) \|_{\R^d} \leq \delta $ implies
$ \| y(t) \|_{\R^d} \leq \epsilon $ for all $  t \geq 0 $.

A good starting point is to consider stability under perturbations in
the space  $ \CC^0_b $
of uniformly  continuous bounded functions 
equipped with the norm $ \| u \|_{\CC^0_b} = \sup_{x \in \R} | u(x) | $.
Since the fixed
point $ u = 0 $ is already unstable
for the ordinary differential equation \reff{a2} it is also
unstable under the dynamics of the partial differential equation
\reff{a3}.
On the other hand, perturbations of $ u = 1 $ in $ \CC^0_b $ can be bounded
by
the maximum principle
using  spatially constant functions, and so
the stability of this equilibrium
follows again from the stability of $ u = 1 $ in the ordinary
differential equation \reff{a2}.

A drawback of this topology is that
the fronts $\vf$ are actually unstable with respect to perturbations
in $ \CC^0_b $. 
In order to see this consider the initial condition
$ w(\xi,0) = \max( v(\xi), \delta ) $. Then, as we see in Fig.\ref{f2}, 
\BENN \sup_{\xi \in \R} | w(\xi,0) - v(\xi) | < \delta , \ {\rm but} \
\lim_{t \rightarrow \infty}
\sup_{\xi \in \R} | w(\xi,t) - v(\xi) | = 1 ~.\end{equation}
However, we shall now argue that
these fronts are stable with respect to spatially
{\em localized} perturbations.
Suppose
that a perturbation is localized near $ \xi = \xi_0>0 $.
Then the perturbation has only a finite time $ t = \xi_0/c $
to grow exponentially
before it is hit by the
bulk of the front,   where  for \reff{a3}
perturbations are damped  exponentially.

\begin{figure}[ht]
\input pra2lab.tex
\epsfig{file=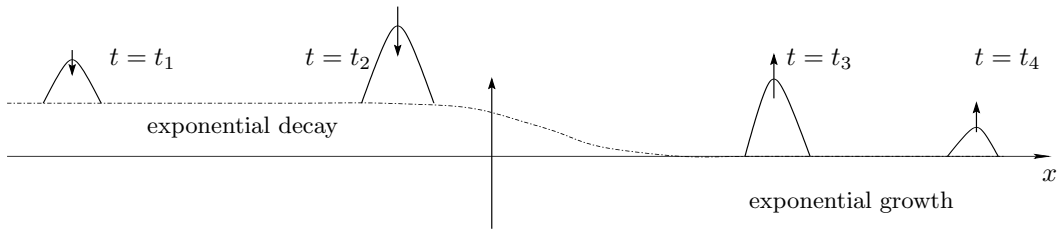, width=14cm,height=3cm, angle=0}
\caption{{ Stability mechanism for  fronts of \reff{a3}. }}
\end{figure}

Next we explain how
this heuristic argument can be made rigorous
with the help of exponentially weighted norms.
In order to do so we consider the stability of the zero solution
$ u \equiv 0 $ for the equation
$ \partial_t u = \partial_{\xi}^2 u + c \partial_{\xi} u + u $.
Let $ w(\xi ) = u(\xi ) e^{\beta \xi} $ with
a $ \beta > 0 $.
Then the  function $ w $ satisfies
\BENN
\partial_t w = \partial_{\xi}^2 w + (c-2\beta) \partial_{\xi} w +
(\beta^2 - \beta c + 1 ) w~.
\end{equation}
If
$ \beta^2 - \beta c + 1 < 0 $ (which we assume henceforth)  
the maximum principle implies the stability of $ w = 0 $ in
$ \CC^0_b $.

However, this reasoning does not carry over to the nonlinear problem
since the product of
two functions $ u_j(x) = w_j(x) e^{-\beta x} $, $ (j=1,2) $
with $ w_j \in \CC^0_b $
leads to $u_1 u_2 e^{\beta x} =
w_1w_2 e^{-2\beta x}e^{\beta x}$, which is in general not bounded for
$x\to-\infty$. 
Therefore, the exponential stability of $ w= 0 $ cannot be used directly.

Only an interplay of both norms,
i.e., the consideration of $ u $ and $ w $ simultaneously, and the
presence of a front
will suffice to show the stability.
We consider thus a front $\vf$ traveling with speed $c$, which is a
solution of \reff{a3} (resp. \reff{moving}).
The deviations $ r = u-\vf  $ and $ w(\xi) =
r(\xi) e^{\beta \xi} $  from the front $ \vf  $ satisfy
\begin{eqnarray*}
\partial_t r &  =
&  \partial_{\xi}^2 r + c \partial_{\xi} r + r - 2 \vf  r - r^2~, \\
\partial_t w &  = &
\partial_{\xi}^2 w + (c-2\beta) \partial_{\xi} w +
(\beta^2 - \beta c + 1 ) w - 2 \vf  w - r w~.
\end{eqnarray*}
At first sight this seems to bring no advantage, but the
term $ -2 \vf  r $ in the equation for $ r $ can be expressed
as
\BENN
- 2 \vf  r = - 2 r + 2 (1-\vf ) r = -2 r + 2 (1-\vf )e^{-\beta \xi}
e^{\beta \xi} r = -2 r + 2 (1-\vf )e^{-\beta \xi} w~.
\end{equation}
Thus, the equation for $ r $ is transformed to:
\BENN
\partial_t r   =
  \partial_{\xi}^2 r + c \partial_{\xi} r -r  + 2 (1-\vf )e^{-\beta \xi} w
  - r^2~.
\end{equation}
The term $  2 (1-\vf )e^{-\beta \xi} $ is {\it bounded} when $\beta ^2
-\beta c +1<0$, since $1- \vf(\xi)= \OO(e^{-\gamma|\xi|}) $ as $\xi\to
-\infty $ with $\gamma>0$ solving $\gamma^2-\gamma c-1=0$.
Because of the term $ -r$ in the equation for $ r $ 
and because we have $ -2 \vf  < 0 $ in the equation for $ w $ 
the linearization of the system $(r,w)$ around
$ (r,w) = (0,0) $ is now {\it exponentially stable} and so this obviously also
holds for the nonlinear system.
\begin{theorem}
Assume $ \beta$ and $ c $ satisfy $  \beta^2 - \beta c + 1 < 0 $.
Then for all $ \epsilon > 0 $ there exist $ C,\delta > 0 $ such that
\BENN
\sup_{\xi \in \R} (| r(\xi,t)| + | w(\xi,t) |)
< C e^{(\max(-1,\beta^2 - \beta c + 1)+\epsilon) t}~,
\end{equation}
provided
\BENN
\sup_{\xi \in \R} | r(\xi,0)\,(1+e^{\beta \xi})|  < \delta~.
\end{equation}
\end{theorem}
The discussion of this equation goes back to \cite{Fi37,KPP37}.
The above result can be found for instance in
\cite{Sa77}.
\begin{remark}
The discussion of the marginally stable case
$ \beta^2 - c \beta + 1 = 0 $ is much more delicate, but covers
the important case of minimal velocity $ |c| = 2 $.
For these solutions
global stability results have been obtained in
\cite{Br83}.
The exact asymptotic  decay of the perturbations
in the sense of the next section has been investigated in \cite{Ga94}.
\end{remark}

\section{The Ginzburg-Landau equation}

\label{sec2gl}
In this section we consider
a model  which comes even closer to the
Taylor-Couette problem, namely the Ginzburg-Landau equation:
\begin{equation}
\label{GL}
\partial_t A =    \partial_x^2 A + A -
 |A|^2A~,
\end{equation}
with $ A=A(x,t)\in \complex $, $ t \geq 0 $, and $ x \in \R $.
This equation has a two-parameter family of nontrivial equilibria
\BENN
\Aeq(x)= \sqrt{1- q^2} e^{i(qx + \vartheta)}~.
\end{equation}
In his pioneering work, Eckhaus \cite{Eck65} showed that the
equilibria with
$ q^2<1/3 $ are linearly stable while those with
$ q^2 > 1/3 $ are linearly unstable.
In contrast to
the equilibrium $ u = 1 $ of the previous problem
the stable equilibria $\Aeq $ are no longer exponentially stable.
Indeed,
the linearization always possesses
continuous spectrum $ (-\infty,0] $ up to the imaginary axis,
i.e.,
at best polynomial decay rates can be expected.

The following discussion can be carried out for all
$ q^2 < 1/3 $. Since our aim is to
explain the basic ideas,
we will restrict ourselves to the stability of the front
$ A(x,t) = B(x-ct) $ connecting
the stable equilibrium
$ \Aeqnull = 1 $ with the unstable equilibrium $ A = 0 $ which
corresponds to the case $q=0$, $\vartheta=0$.

The real-valued front $B=\Bfront(\xi) $ satisfies the ordinary differential equation
$ B'' + c B' + B - B^3 = 0 $.
It
can be found by
analyzing this second order
ordinary differential equation
in the $ (B,B') $--phase plane. There exists a heteroclinic connection
between $ (B,B') = (1,0) $ and $ (B,B') = (0,0) $, where the latter
fixed point has real eigenvalues for $ |c| \geq 2 $.
As a consequence, when $ |c | \geq 2 $ the fronts satisfy
$  B(\xi) \in (0,1) $.

We introduce the deviation $ v $ from the front by writing
$ A(x,t) = \Bfront(x-ct)+ v(x,t) $, and 
if $A$ satisfies \reff{GL}
then $ v $ solves the equation
\begin{equation} \label{bes1}
\partial_t v = \partial_x^2 v  + v - \Bfront^2 (2 v+ \overline{v})
- v^2  {\Bfront} - 2 |v|^2 \Bfront - v|v|^2~.
\end{equation}
Again,
the stability of the front will follow
only by an interplay of the usual  norm and the
spatially exponentially weighted norm.
Therefore, we introduce $ w(\xi) = v(\xi+ct) e^{\beta \xi} $
and the equation for $w$ is
\begin{eqnarray*}
\partial_t w &  = &  \partial_{\xi}^2 w +
 (c-2 \beta ) \partial_{\xi} w + (\beta^2 - c  \beta +1) w \\
&&
- \Bfront^2 (2 w+ \overline{w})
- \Bfront vw - 2\Bfront  \overline{v} w - |v|^2 w ~,
\end{eqnarray*}
where $ \xi = x-ct $.
When $ \beta^2 - c  \beta +1 = -2 \gamma < 0$, the maximum principle 
applied to the system written in polar coordinates $w=|w|e^{i\phi}$ 
implies
the exponential stability of $ w = 0 $ and we get a bound
$ w = \OO(\exp(-\gamma t)) $ as $ t \rightarrow \infty $ provided 
$ v $ stays bounded. To exploit the effect of the bulk, we write 
the difference between  $ \Bfront  $ and the stable equilibrium $ \Aeqnull =
1 $ as
$ g=\Bfront-1 $.
Using $g$, \reff{bes1} can be written as
\begin{eqnarray*}
\partial_t  v & = & \partial_x^2 v +  v -(2 v+ \overline{v})
- v^2 - 2 |v|^2- v|v|^2 \\
&&
- (2g+g^2) (2 v+ \overline{v})
-  v^2  g - 2 |v|^2  g~.
\end{eqnarray*}
All  terms in the second line vanish at some exponential rate
as $ x-ct \rightarrow - \infty $.
We use this as above by expressing $ v $ in terms of $ w $,
for instance,
$
g v = (e^{-\beta \xi} g)(e^{\beta \xi} v)= (e^{-\beta \xi} g) w$,
and so
\BENN
\partial_t  v  =  \partial_x^2 v + v -(2 v+ \overline{v})
- v^2 - 2 |v|^2- v|v|^2 + \OO(w)~.
\end{equation}
Writing $A$ in polar coordinates
$ A = (1 + r) e^{i \phi} $ we then find
\begin{eqnarray}
\partial_t r &  =
&  \partial_x^2 r   - 2 r - (\partial_x\phi)^2(1+r)
-3r^2 -r^3 + \OO(w)~,\\
\partial_t \phi & = &  \partial_x^2\phi
+ 2 \frac{(\partial_x\phi)(\partial_x r)}{1+r} + \OO(w)~.
\label{bayn1}
\end{eqnarray}
Linearizing this system, ignoring
the terms $ \OO(w) $ which decay at some exponential rate, gives
\BENN
 \partial_t r  =
 \partial_x^2 r  - 2 r
\ \ \ {\rm and} \ \ \
\partial_t \phi =\partial_x^2\phi~.
\end{equation}
This shows that the variable $  r $ decays with some exponential rate.
However, the   variable $\phi $ shows some diffusive behavior.
Therefore,
the variable $ r $
can be expressed asymptotically by $ \phi $,
that is,
$ r = -(\partial_x \phi)^2/2 + {\rm h.o.t.}\, $.
Inserting this into the equation for $ \phi $ yields
\BENN
\partial_t \phi = \partial_x^2  \phi - (\partial_x \phi)
\partial_x(\partial_x
\phi)^2
+ {\rm h.o.t.}~. \end{equation}
We explain now the arguments which are needed
to analyze this equation.
In order to do so we go back to the linearized system.
In Fourier space $ \phi  $ satisfies
$
\partial_t \tilde{ \phi}(k,t) = - k^2 \tilde{ \phi}(k,t)$,
that is $ \tilde{ \phi}(k,t) = \exp(-k^2 t)
\tilde{ \phi}(k,0) $.
Thus, the Fourier modes of  $ \tilde{ \phi} $ are concentrated around
$ k=0 $ as time evolves and so 
\BENN 
\tilde{ \phi}(k/\sqrt{t},t) = \exp(-k^2 )
\tilde{ \phi}(k/\sqrt{t},0) = \exp(-k^2 )\tilde{ \phi}(0,0) + \OO(1/\sqrt{t})~,
\end{equation}
for 
sufficiently smooth $ \tilde{ \phi}(k,0) $.
Since  smoothness in Fourier space corresponds to decay rates
for $ |x| \rightarrow \infty $
this explains the
well known fact that solutions to
spatially localized initial conditions  decay to zero in
a universal manner,
\BENN
\phi(x,t) = \frac{\phistar}{\sqrt{t}} \exp(\frac{-x^2}{4t}) +
\OO(t^{-1})~, 
\end{equation}
with some constant  $ \phistar \in \real $ depending only on the
initial conditions.
As a consequence, we have
\BENN
\partial_t \phi = \OO(t^{-3/2}), \
\partial^2_x \phi = \OO(t^{-3/2})~, \end{equation}
but under these asymptotics,
\BENN
- (\partial_x \phi)\partial_x (\partial_x \phi)^2
= \OO(t^{-7/2})~.
\end{equation}
Therefore, all nonlinear terms in the equations for $(r,\phi)$
vanish much faster
than those which are part of the linear diffusion equation.
In general,
nonlinear terms $ \phi^{p_1}(\partial_x \phi)^{p_2}
(\partial_x^2 \phi)^{p_3} $  are therefore called irrelevant if
$ p_1 + 2 p_2 + 3 p_3 > 3 $.

With the help of renormalization theory these heuristic arguments
have been made rigorous and it has been shown that the
nonlinear system possesses the same asymptotics \cite{CEE92,BK92}.
Thus spatially localized perturbations of the equilibrium  $ \Aeqnull=1 $
vanish asymptotically like solutions of a linear diffusion problem.

{\bf Notation.}
Let $ \| u \|_{\L^2}^2 = \int |u(x)|^2 dx $. We define
the usual Sobolev norm by $ \| u \|_{{\rm H}^2} = \| u \|_{\L^2} + \|
\partial_x  u \|_{\L^2}+ \|
\partial_x^2  u \|_{\L^2} $. Moreover, we set
$ \| u \|_{\Hhh} = \| u \rho  \|_{{\rm H}^2} $, where
$ \rho(x) = (1+x^2)^{1/2} $.
\begin{theorem} \label{glasy}
There exist positive constants $ \delta $,
and $C$,
such that for all $v_0$ with $ \|v_0\|_{\Hhh} < \delta  $
the following holds. Let
$ A = \Aeqnull+v$
be the solution 
of the
Ginzburg-Landau equation \reff{a3}
with initial condition $x\mapsto \Aeqnull(x)+v_0(x)=1+v_0(x)$.
Then $v$ exists for all times $t\geq 0$ and there exists a
constant $v_*\in\real$ depending only on the initial condition $v_0$
such that for all $t\ge1$ one has 
\BENN \sup_{x \in \R}
|  v(x,t) - \frac{\vlim }{\sqrt{t}}
\exp(\frac{-x^2}{4  {t}}) | \leq
 \frac{C}{t}~.
\end{equation}
\end{theorem}
Furthermore,
spatially localized perturbations of the front $ A(x,t) =\Bfront(x-ct) $
vanish asymptotically as a solution of a linear diffusion equation
\cite{BK94,EW94}.
\begin{theorem} \label{glasy1}
Let $\Bfront$ be a front with velocity $ c > 2 $.
Let $ \beta>0 $ be such that $ \beta^2 - c \beta + 1 = - 2
\gamma < 0 $. 
Then
there exist positive constants $ \delta $
and $C$
such that 
for all $v_0$ with $ \|x\mapsto v_0(x)(1+e^{\beta x})\|_{\Hhh} < \delta  $
the following holds.
Let
$ A(x,t) =  \Bfront(x-c t) + v(x,t) $ be the solution
of the
Ginzburg-Landau equation \reff{a3}
with initial condition $A_0(x)= \Bfront(x) + v_0(x) $.
Then $v$ exists for all $t\ge0$ and there is a constant $\vlim$
depending only on $v_0$ such that for all $t\ge 1$
one has
\BENN \sup_{x \in \R}
|  v(x,t) - \frac{\vlim }{\sqrt{t}}
\exp(\frac{-x^2}{4  {t}}) | \leq
\frac{C}{t}~,
\end{equation}
and \BENN \sup_{\xi \in \R}
| v(\xi+ct,t)e^{\beta \xi} | \leq C e^{-\gamma t}~.
\end{equation}
\end{theorem}
The case of minimal velocity $ c = 2 $,
which is more delicate, has been handled in \cite{BK94,EW94} with
norms which take more details
of $ \Bfront $ into account. 
\begin{remark}
These fronts are closely related to the modulated fronts
of the Taylor-Couette problem.
Indeed, using the multiple scaling ansatz
\citel{NW69,dPES71},
\BENN
U = {\psi}_{\eps,A} + \OO(\epsilon^2),
\quad {\rm where} \
{\psi}_{\eps,A}(x,t)=\eps \ A(\epsilon x,\epsilon^2 t) \
\hhU(x) e^{i \kc  x} +
{\rm c.c.}~,
\end{equation}
one sees that  the Taylor-Couette problem is approximated
by the Ginzburg-Landau equation
for the $A$.
Mathematical theorems which show exact relations between
the Ginzburg-Landau equation and the Taylor-Couette system
can be found in \cite{CE90b,vH91,KSM92,Schn94a,Schn94b,Schn98}.
\end{remark}

\section{The Taylor-Couette problem}

In line with the discussion of the previous sections
the stability proof for
the modulated
fronts $\Umf (x- ct,x,z)$
mainly consists of two parts:
i) The introduction of an exponential weight to stabilize
the unstable part ahead of the front and
ii) A stability analysis
of the stationary solution $ \Utv
$ behind the front.
As  above we will restrict ourselves in this paper to the case $ q = 0 $
which lies in the Eckhaus-stable region.

Note that for this problem the stability analysis of the equilibrium
is by no means obvious, but can be found in \cite{Schn98}.
The
equilibrium $\Utvnull$ of Eqs.\reff{ns1}--\reff{ns4} possesses
continuous spectrum up to the imaginary axis just like
$\Aeqnull$ of Eq.\reff{GL}. 
Therefore, one can expect at best polynomial decay in time.
Furthermore, it has been shown that the
 linearly stable Taylor vortices are also nonlinearly
stable with respect to small spatially
localized perturbations and that the perturbations vanish
in a universal way like a solution of a
linear diffusion equation.
\begin{theorem}
\cite{Schn98}. Let  $\epsilon >0$ sufficiently small.
Then there exist   $ \delta>0 $, and constants $C_1>0$, $C_2 $
such that for
all $ V_0 $ with
$ \| V_0  \|_{\Hhh} < \delta $  the following holds.
Let $ U = \Utvnull +
 V $ be the solution of
the system
\reff{ns1}--\reff{ns4} with initial condition $(x,z)\mapsto\Utvnull(x,z) +
 V_0(x,z) $.
Then $ V $
exists for all times $ t \geq 0 $,
and there exists a constant
$ \Vlim  \in \R $
depending only on the initial condition $V_0$
 such that for all $ t \geq 1 $ one has
\BENN
 \sup_{(x,z) \in \R \times \Sigma}
 | V(x,z,t) -  \frac{1}{\sqrt{ t}}
 \Vlim 
\exp(\frac{-x^2}{4 C_{1}t}) \partial_x \Utvnull(x,z) |
 \leq
\frac{C_2} { t^{3/4}}~.
\end{equation}
\end{theorem}
The corresponding result for the Swift-Hohenberg equation can be found in
\cite{Schn96,EWW97}.

After the stability analysis of the part behind the front
it remains to stabilize
the zero solution ahead of the front.
We again use the interplay of two norms.

We introduce $ W $ by $ W(\xi,z,t) = V(\xi+ct,z,t) e^{\beta \xi} $.
If we insert this into
the linearization around the Couette flow we obtain an estimate
for the variable $ W $, namely
\BENN
\| W(\cdot,\cdot,t) \|_{{\rm H}^2} \leq C e^{-\rho(c,\beta,\epsilon) t}
\| W(\cdot,\cdot,0) \|_{{\rm H}^2}~.
\end{equation}
For given $ \epsilon > 0 $, sufficiently small,
 there is a minimal velocity
$ c_0 > 0 $ such that for all $ c > c_0 $ there are $ \beta_l $
and $ \beta_u $ such that for all $ \beta \in (\beta_l,\beta_u) $ one has
$ \rho(c,\beta,\epsilon) > 0 $.
Given such an $ \epsilon > 0 $ and a $ c > c_0 $ we choose now a
$ \beta $ such that
\BENN
\rho(c,\beta,\epsilon) = 2 \gamma > 0~.
\end{equation}
Then we have:
\begin{theorem}
\label{thhaupt}
Let $\epsilon >0$ sufficiently small.
Then there exist a $ \delta>0 $, and constants $C_1>0$, $C_2 $
such that for
all $ V_0 $ with
$ \| (x,z)\mapsto V_0(x,z) (1+e^{\beta x}) \|_{\Hhh} < \delta $
the following holds.
Let $ U(x,z,t) = \Umfnn(x-ct,x,z) +
 V(x,z,t) $ be a solution of
the system
\reff{ns1}--\reff{ns4}  with initial condition $(x,z)\mapsto
\Umfnn(x,x,z) +
 V_0(x,z) $.
Then $ V $
exists for all times $ t \geq 0 $,
and there exists a constant
$ \Vlim  \in \R $
depending only on the initial condition $V_0$ 
 such that for all $ t \geq 1 $ one has
\begin{equation}\label{full}
 \sup_{(x,z)  \in \R \times \Sigma}
 | V(x,z,t) -  \frac{1}{\sqrt{ t}}
 \Vlim 
\exp(\frac{-x^2}{4 C_{1}t}) \partial_x \Utvnull(x,z) |
 \leq
\frac{C_2}{  t^{3/4}}~,
\end{equation}
and
\BENN
 \sup_{(\xi,z)  \in \R \times \Sigma}
 | V(\xi+ct,z,t)e^{\beta \xi}  |
 \leq
C_2  e^{-\gamma t}~.
\end{equation}
\end{theorem}
{\bf Sketch of proof.}
We proceed as in the previous sections.
First we write \reff{ns1}--\reff{ns4} as a dynamical system
in the space of divergence free vector fields \cite{He81}.
Then we consider the system for $ V $ and the exponentially weighted
variable $ W $.
The proof is exactly the same as in \cite{ES00} given
for the fronts in the Swift-Hohenberg equation.
The analysis behind the front and the complete functional analytic
set-up can be found in \cite{Schn98}. Finally, 
in \cite{Schn99}, one finds the estimate
showing that the Taylor-Couette problem
is approximated by the Ginzburg-Landau equation. This is used ahead
of the front.
\qed
\begin{remark}
One can interpret \reff{full} as saying that the perturbations decay
faster near the extrema of $U_{\rm TV}$. This can also be understood
in the sense that at large amplitude of the underlying periodic
pattern (i.e., where $\partial_x {U}_{\rm TV}(x,z) =0$),
the restoring force is stronger.
\end{remark}
\begin{remark}
The choice  of a sufficiently small
$ \epsilon > 0 $
actually allows
to
prove the stability of {\it all} fronts which are predicted to be stable
by the associated amplitude equation (the Ginzburg-Landau equation)
since
\BENN 
\lim_{\epsilon \rightarrow 0}
\epsilon^{-2} \varrho(\epsilon c_B,\epsilon \beta_A,\epsilon) = 
\varrho_A(c_B,\beta_A)~,
\end{equation}
where $ \varrho_A(c_B,\beta_A) $ is the stability condition
of the associated amplitude equation.
\end{remark}
\begin{remark}
Note that the Taylor-Couette problem is structured by the periodic
background of the stationary solution. In such a setting it is natural
to
study perturbations (which need not be periodic) in the so-called
Bloch wave representation (see e.g.~\cite{RS72}).
This means that one writes
\BENN
u(x)=\int_{-1/2}^{1/2} e^  {i\ell x} \hat u(\ell,x)~,
\end{equation}
where
\BENN
\hat u(\ell,x)= \sum_{n\in{\integer }} e^{inx }\tilde u(n+\ell)~,
\end{equation}
with $\tilde u$ the Fourier transform of $u$.
Since the linearized problem for the Taylor vortices is periodic, the
corresponding linear operator is diagonal in the $\ell$ and thus the
Bloch representation simplifies the analysis.
Nonlinear terms can be written in Bloch space with the help of
convolution integrals. Decay for $ |x| \rightarrow \infty $ corresponds
in Fourier and Bloch space to smoothness with respect to the
Fourier wave number $  k $ and Bloch wave number $ \ell $, respectively.
As a consequence, the important relation
$ \partial_x \cdot = \OO(t^{-1/2}) $ for the irrelevance of terms corresponds
in Bloch space to $  \ell \cdot = \OO(t^{-1/2}) $.
\end{remark}
\begin{remark}
The diffusive behavior of solutions
close to the Taylor vortices can be understood as follows:
The eigenfunctions of the linearization of the Taylor vortices are
given by Bloch waves $ e^{i\ell x} v_n(\ell, x,z) $ with
$ v_n(\ell, x,z) =v_n(\ell, x + 2 \pi /\kc  ,z) $ having the same periodicity
as the Taylor vortices. For each fixed Bloch wave number $ \ell $ there is
discrete spectrum, i.e., $ n \in {\natural } $.
Therefore, there are smooth curves $ \mu_n(\ell) $ of eigenvalues
over the  Bloch wave numbers.

\begin{figure}[ht]
\input pra4lab.tex
\epsfig{file=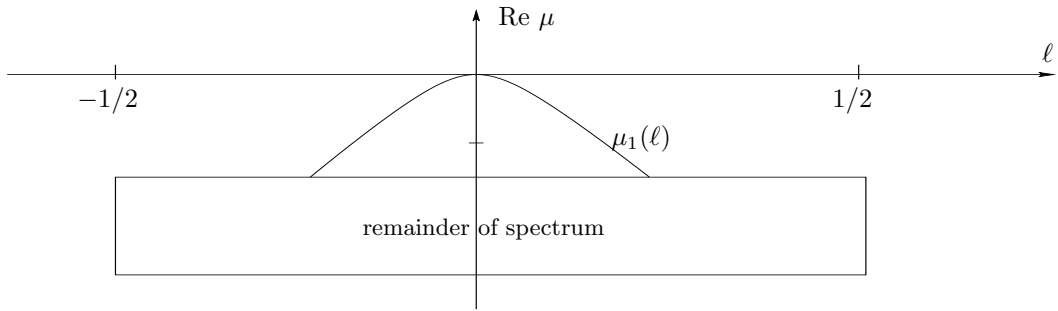, width=14cm,height=4cm, angle=0}
\caption{{ Spectrum of the linearization around the Taylor vortices
as a function over the Bloch wave numbers.}}
\end{figure}

There is one curve, $ \mu_1 $, coming up to zero.
This curve has approximately the form of a parabola,
i.e., $ \mu_1(\ell) = - c_1 \ell^2 + \OO(\ell^3) $ with $ c_1 > 0 $
in the Eckhaus stable region.
Thus, the solution $ \hat{v}_1 $ rescaled in Bloch space
satisfies
\BENN \lim_{t \rightarrow \infty} \hat{v}_1(\ell/\sqrt{t},t) = v_* e^{-c_1 \ell^2}~,
\end{equation}
for smooth initial conditions in Bloch space,
and shows the same asymptotic behavior as the linear
diffusion equation $ \partial_t v_1 = c_1 \partial_x^2 v_1 $ for spatially
localized initial conditions.
\end{remark}
\begin{remark}
As for the case of the Ginzburg-Landau equation we have,
for the perturbations of the Taylor vortices, asymptotically irrelevant
nonlinearities. This is not obvious and has been proved in
\cite{Schn98} by deriving an effective equation
for the variable corresponding to the curve of critical eigenvalues
$ \mu_1 $ and proving that the coefficients
in front of the quadratic and cubic nonlinear terms vanish up to a certain
order  as the Bloch wave number $
\ell $ goes to zero.
These facts are a reflection of the translation invariance and of the
existence of a circle of fixed points for the stationary problem.
\end{remark}
\begin{remark}
The nonlinear stability of so-called critical fronts (moving at
the minimal possible speed for which they are linearly stable) remains open. 
See \cite{CE87} for the linear stability analysis of the fronts in the 
Swift-Hohenberg equation. 
Achieving
this aim seems to be a necessary step in solving the long-standing
problem of ``front selection'' \cite{DL83}, in a case where the maximum
principle \cite{AW78} is not available.
\end{remark}

\end{document}